\documentclass[twocolumn,amssymb,nofootinbib,nobibnotes,aps,prd]{revtex4}

\usepackage{graphicx}
\usepackage[dvips]{color}
\usepackage{amsmath,amsthm,amssymb}

\newcommand{\be}{\begin{equation}}
\newcommand{\ee}{\end{equation}}
\newcommand{\bd}{\begin{displaymath}}
\newcommand{\ed}{\end{displaymath}}
\newcommand{\ba}{\begin{eqnarray}}
\newcommand{\ea}{\end{eqnarray}}

\def\gr{$\gamma$-ray}

\def\ap{\approx}

\def\e{{\rm e}}

\def\lsim{\raise0.3ex\hbox{$\;<$\kern-0.75em\raise-1.1ex\hbox{$\sim\;$}}}
\def\gsim{\raise0.3ex\hbox{$\;>$\kern-0.75em\raise-1.1ex\hbox{$\sim\;$}}}
\def\eps{\varepsilon}
\def\theta{\vartheta}

\def\Rm{{\cal R}_{\rm max}}

\begin{document}

\title{Unified model for cosmic rays above $10^{17}$\,eV and the diffuse
gamma-ray and neutrino backgrounds}

\author{G.~Giacinti$^{1,2}$}
\author{M.~Kachelrie\ss$^{3}$}
\author{O.~Kalashev$^{4}$}
\author{A.~Neronov$^{5}$}
\author{D.~V.~Semikoz$^{6}$}

\affiliation{$^{1}$Max-Planck-Institut f\"ur Kernphysik, Postfach 10 39 80, 69029 Heidelberg, Germany}
\affiliation{$^{2}$University of Oxford, Clarendon Laboratory, Oxford OX1 3PU, United Kingdom}
\affiliation{$^{3}$Institutt for fysikk, NTNU, Trondheim, Norway}
\affiliation{$^{4}$Institute for Nuclear Research of the Russian Academy of Sciences, Moscow, Russia}
\affiliation{$^{5}$ISDC, Astronomy Department, University of Geneva, Ch.~d'Ecogia 16, Versoix 1290, Switzerland}
\affiliation{$^{6}$AstroParticle and Cosmology (APC), 10 rue Alice Domon et L\'eonie Duquet, F-75205 Paris Cedex 13, France}

\begin{abstract}
We investigate how the extragalactic proton component derived within the 
``escape model''  can be explained by astrophysical sources. We consider as 
possible cosmic ray (CR) sources normal/starburst galaxies and radio-loud 
active galactic nuclei (AGN). We find that the contribution to the total 
extragalactic proton flux from normal and starburst galaxies is only 
subdominant and does not fit the spectral shape deduced in the escape model. 
In the case of radio-loud AGN, we show that the 
complete extragalactic proton spectrum can be explained by a single source 
population, BL Lac/FR I, for any of the potential acceleration sites in these
sources. We calculate the diffuse neutrino and \gr\ fluxes produced 
by these CR protons interacting with gas inside their sources.
For a spectral slope of CRs close to $\alpha=2.1-2.2$
as suggested by shock acceleration, we find that these 
UHECR sources contribute the dominant fraction of both the isotropic \gr\ 
background and of the extragalactic part of the astrophysical neutrino signal 
observed by IceCube. 
\end{abstract}


\maketitle



\section{Introduction}
\label{Introduction}

The search for the sources of ultrahigh energy cosmic rays (UHECR) 
and for an understanding of their acceleration mechanism is one
of the important challenges in astroparticle physics. It has been hoped for
that CRs at the highest energies would be only weakly deflected in magnetic
fields and hence sources could be identified by usual astronomical methods.
Both the Pierre Auger Observatory (PAO) and the Telescope Array (TA) do
observe anisotropies in the arrival directions of 
UHECRs~\cite{PierreAuger:2014yba,Abbasi:2014lda} and a
hot spot observed by TA has also a large statistical significance. 
However, these anisotropies extend over medium angular scales, 
similar as found in Ref.~\cite{Kachelriess:2005uf} 
for previous experimental data, and no successful correlation of 
UHECRs with potential astrophysical sources has been achieved yet. Meanwhile, 
there has been steady progress in other areas: The all-particle CR spectrum 
has been measured precisely, and data on the primary composition have become 
available. The Auger collaboration provided in~\cite{Aab:2014aea} constraints 
on the fraction of four different elemental groups above $6\times 10^{17}$\,eV,
while the KASCADE-Grande experiment covered with its composition measurements
energies up to $2\times 10^{17}$\,eV~\cite{dataKG}. Although the derived 
composition depends 
on the hadronic interaction models used for the analysis, the following 
qualitative conclusions can be drawn: First, the proton fraction amounts to
$\sim 40$--60\% in the energy range between $7\times 10^{17}$\,eV and 
$7\times 10^{18}$\,eV and decreases afterwards, while the fraction of 
intermediate nuclei (He, N) increases. Second, the iron fraction in the 
energy range between $7\times 10^{17}$\,eV and $2\times 10^{19}$\,eV is 
limited by $\lsim $15--20\% and 
its central value is consistent with zero. Thus the Galactic contribution 
to the observed CR spectrum has to die out around $7\times 10^{17}$\,eV, 
unless an additional sub-dominant and heavy Galactic component remains. 
Both light and intermediate elements above this energy cannot have a Galactic 
origin because their anisotropy would otherwise overshoot the upper limits 
set by Auger on the CR dipole anisotropy, see Ref.~\cite{dipole}.

These results provide a strong constraint on models for the
transition between the Galactic and the extragalactic CR component.
In particular, they exclude the dip model~\cite{dip} which requires a proton 
fraction $\gsim 90\%$. In other models, the ankle at $E\ap {\rm a~few} 
\times 10^{18}$\,eV has been identified with 
the transition between Galactic and extragalactic CRs~\cite{ankle1,ankle2}.
However, such a high value of the transition energy exaggerates the 
acceleration problem of Galactic CR sources and contradicts the low iron
fraction determined by Auger.
Alternatively, it has been suggested that two populations
of extragalactic CR sources exist, one dominating below and one above the 
ankle~\cite{Aloisio:2009sj}. Since there exists no convincing model for
these two source classes and their properties, a more economical explanation
based on a single extragalactic source type is desirable.

Several recent studies tried to explain the measured spectrum and 
the composition both above and below the ankle using models with a mixed 
composition~\cite{Globus:2015xga,Unger:2015laa,Taylor:2015rla}. Two of them,
Refs.~\cite{Globus:2015xga,Unger:2015laa}, explain the 
light composition below the ankle by  photo-disintegration  in the  CR
sources, while at higher energies the initial mixed composition survives. 

In  Ref.~\cite{Unger:2015laa}, the energy spectra of nuclei in the source are 
assumed to follow a $1/E$ power-law. As a result, the extragalactic sources 
do not contribute at $E\lsim 3\times 10^{17}$ eV.
In Ref.~\cite{Globus:2015xga}, the original index of nuclei spectra is around 
$2.1-2.2$, 
as required by acceleration models, but all low energy nuclei 
are photo-disintegrated in the sources. As a result, 
the spectra of nuclei leaving the source are close to a 
$1/E$ power-law, while the proton spectrum follows the original acceleration 
spectrum because of a decaying neutron component which escapes from the source. 
These models predict a contribution of extragalactic 
sources to the proton spectrum in the energy range measured by KASCADE-Grande. 

A disadvantage of this type of models is that they predict a negligible 
contribution from extragalactic CR sources to the observed IceCube neutrino 
signal. Thus the extragalactic neutrino flux is 
disconnected from the extragalactic CR flux, requiring that extragalactic 
neutrinos are produced in ``hidden sources'' with a large interaction depth 
for protons.
In both models, sources with strongly positive evolution give the major 
contibution to the UHECR flux.  In contrast, the authors of Ref.~\cite{Taylor:2015rla} study sources with negative evolution in order to 
explain the composition measured by Auger with softer fluxes $\propto 1/E^2$ 
without invoking photo-disintegration.

In this work, we investigate which CR source classes can explain the 
extragalactic proton component derived within the escape model~\cite{PI,PII} 
and, at the same time, can give a significant
contribution to the isotropic \gr\ background (IGRB) measured by Fermi-LAT~\cite{Fermi} 
and to the astrophysical neutrino signal observed by IceCube~\cite{Ice}. 
These neutrinos have been suggested to have either a Galactic 
(e.g.~\cite{Neronov:2013lza,Taylor:2014hya,Fox:2013oza}) 
and/or an extragalactic origin (e.g.~\cite{Ahlers:2014ioa}),
and may have a hadronuclear origin~\cite{Murase:2013rfa,Ahlers:2013xia,Padovani:2015mba}.
 Amongst others, radio-loud AGNs and hosting galaxy clusters have being proposed as possible sources~\cite{Murase:2013rfa}, 
and a connection to $10^{17.5-18.5}$\,eV CRs was suggested and investigated in~\cite{Murase:2008yt}. 
Secondary neutrinos and \gr s from UHECR propagation in clusters
were calculated in Ref.~\cite{Kotera:2009ms}.

We consider here as possible 
CR sources normal/starburst galaxies and radio-loud active galactic nuclei 
(AGN). This choice is motivated by the fact that these sources might give 
the dominant contribution  to the IGRB at low and high energies, 
respectively. In particular, it has been suggested in 
Refs.~\cite{Stecker:1996ma} 
that blazars can contribute up to 100\% to the IGRB. 
As production mechanism of the secondary \gr\ and neutrino
fluxes, we use CR interactions with gas in their sources.
We find that normal and star-forming galaxies
can explain neither the spectral shape nor the magnitude of the 
derived extragalactic proton flux. In the case of  radio-loud AGNs, we 
show that the complete extragalactic proton 
spectrum can be explained by a single source population. We calculate also 
the diffuse neutrino and \gr\ fluxes produced by these CR protons interacting 
with gas.  For a spectral 
slope of CRs close to $\alpha_p=2.1-2.2$ as suggested by shock acceleration, 
we find that these UHECR sources can contribute the dominant fraction to 
the isotropic \gr\ background (IGRB), and the major contribution to the 
extragalactic part of the astrophysical neutrino signal observed by IceCube.


\section{Methodology}
\label{M}

\subsection{Extragalactic CR proton flux}

\begin{figure}
  \includegraphics[width=0.35\textwidth,angle=270]{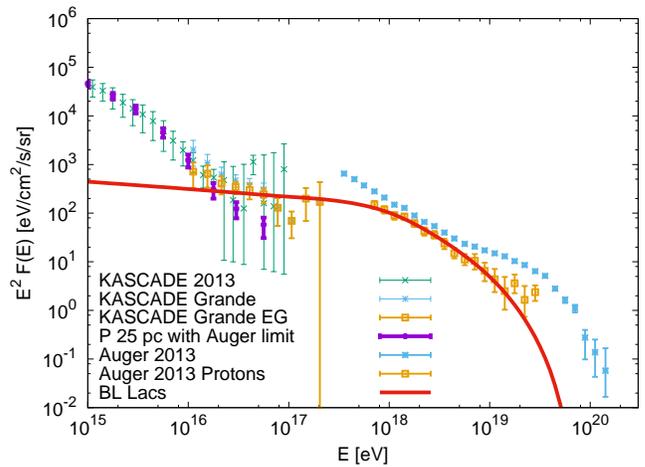}
  \caption{Extragalactic proton flux deduced in Ref.~\cite{PII} within
  the escape model from  KASCADE-Grande  and Auger  data (brown errorbars)
  together with the original KASCADE and KASCADE-Grande proton data and
  the total  CR flux from Auger. Also shown is the  predicted proton flux
  from BL Lacs (red line).
  \label{Fig_exprot}}
\end{figure}

We summarize first how the extragalactic CR proton flux in the 
escape model was derived in Ref.~\cite{PII}.
In a first step,  we derived the Galactic all-particle CR flux summing up all 
CR groups obtained for the maximal rigidity $\Rm= 10^{17}$\,V and 
accounting for the Auger iron constraint. Then we obtained the total 
extragalactic flux by subtracting the predicted total Galactic flux from the 
measured total CR flux. The extragalactic proton flux  observed by Auger
followed applying the composition measurement~\cite{Aab:2014aea} where we 
chose the results obtained using the EPOS-LHC simulation. 
Finally, we deduced the contribution of extragalactic
protons to the observed proton flux by KASCADE and KASCADE-Grande,
subtracting from the proton fluxes given in Ref.~\cite{dataKG} the 
prediction of the escape model: Since the predicted Galactic proton flux lies 
for energies above $E\gsim 3\times 10^{16}$\,eV below the measured one,
the difference has to be accounted for by extragalactic protons. 
We show in Fig.~\ref{Fig_exprot} the deduced extragalactic proton 
fluxes from KASKADE-Grande and Auger data  in the escape model with
brown errorbars. 
Combining the  KASCADE, KASCADE-Grande and Auger data suggests that 
the slope of the extragalactic proton energy spectrum is flat
at low energies, $E\lsim 10^{18}$\,eV,  consistent with 
$\alpha_p\sim 2.2$, and softens to $\alpha_p\sim 3$ at 
higher energies, $E\gsim 10^{18}$\,eV, cf.\ Fig.~\ref{fig:spectrum}.

\subsection{CR interactions with gas and photons}
\label{CRinterac}

We summarize now how we calculate the interactions of CRs with gas and
the extragalactic background light (EBL). We split the propagation in two
parts: The first one includes the propagation in the source, the host galaxy and
galaxy cluster where we assume that proton interactions with gas are dominant. 
The spectrum of exiting particles is then used in the second step as an 
``effective source spectrum'' from which we calculate the resulting 
diffuse flux taking into account the distribution $\rho(z,L)$ of sources 
as well as the interaction of protons, electron and photon with the 
EBL and the CMB. For both steps, we use the open source 
code~\cite{Kalashev:2014xna} which solves the corresponding kinetic 
equations in one dimension. We employ the baseline EBL 
model of Ref.~\cite{EBL}.

As input for the first step, we require the energy dependent grammage $X(E)$ 
and the proton injection spectrum $dN_{CR}/dE$. Starting from the 
injection spectrum of protons, we simulate their propagation and
their interaction  to obtain the spectra of protons and secondary particles 
leaving the ``effective source''. We neglect all interactions except $pp$ 
interactions in the ``effective source''. This assumption is not satisfied 
for some UHECR acceleration sites in the vicinity of the AGN, such as notably acceleration close to the accretion disk 
where the intense radiation field would make $p\gamma$ interactions dominate 
over $pp$. However, for acceleration at the polar caps, the radiation field is sufficiently low 
for $pp$ to dominate. We also assume that all secondaries 
escape freely, except electrons. The fate of electrons depends on the strength 
of the radiation and magnetic fields
and on the source size. In extended sources as galaxies with relatively
small magnetic fields, they lose all their energy via synchrotron and
inverse Compton radiation. 
For simplicity, we neglect pair production by photons 
inside the source, since the following cascade development outside the source 
leads to a universal spectrum.

The code~\cite{Kalashev:2014xna} has been extended implementing $pp$
interactions as follows: 
The inelastic cross sections $\sigma_{\rm inel}$ of CR nuclei on gas were 
calculated with QGSJET-II-04~\cite{qgs}.  For the spectrum of secondary 
photons and neutrinos produced in $pp$ interactions differential cross 
sections tabulated from QGSJET-II-04 were used~\cite{ppfrag}. Secondaries 
from heavier elements in the CR flux are suppressed. For this reason, 
we will only need to consider the extragalactic CR proton flux in the following, when trying 
to fit CR, gamma-ray and neutrino fluxes altogether. 
The contributions from CR nuclei are included adding a nuclear enhancement factor $\eps_M$. 
Similarly, the helium component of the interstellar medium was accounted for. Combining both 
effects, we set $\eps_M= 2.0$~\cite{enh}. 
In order to take into account properly the energy dependence of the 
grammage, while still using kinetic equations in a one-dimensional framework, 
we include in the $pp$ interaction rates $R(E)$ the energy dependent 
grammage $X(E)$. 
In the case of radio-loud AGNs, we have used for the grammage the simple 
parameterization $X(E)\propto E^{-1/3}$ expected for a turbulent magnetic field 
with a Kolmogorov spectrum. The normalization was fixed by setting the 
interaction depth $\tau_{pp}=1$ at a reference energy 
$E_{\rm esc}$, which is the only free model parameter in this case. 
In contrast, for star-forming galaxies, we have used the grammage 
derived within the ecape model, see next section.

\section{Star-forming galaxies}
\label{StarformingGal}

The escape model developed in Refs.~\cite{PI,PII}  provides
an excellent description of the measured fluxes of CR nuclei from
below the knee to $\sim 10^{18}$\,eV. The model predicts an early
transition from Galactic to extragalactic CRs. Natural candidate 
sources for (part of) the extragalactic flux are all other star-forming 
galaxies (i.e.\ normal spiral galaxies, starburst galaxies and star-forming 
AGNs). 
Moreover, starburst galaxies are expected to give a major contribution to 
the IGRB, and they have also been considered as possible sources of 
high-energy neutrinos, assuming they are CR calorimeters --see amongst others 
Refs.~\cite{Loeb:2006tw,Liu:2013wia,Tamborra:2014xia,Anchordoqui:2014yva}. 
Therefore, we first examine how large the contribution from 
star-forming (SF) galaxies to these fluxes is.

\subsection{CR flux from a single galaxy}
\label{single}

We apply the escape model to other normal spiral galaxies. In 
Refs.~\cite{PI,PII}, we argued that a Kolmogorov spectrum for the
turbulent galactic magnetic field together with
a $1/E^{2.2}$ power-law injection spectrum provides an explanation
of the knee and of all CR composition measurements between $10^{14}$\,eV
and $10^{18}$\,eV. The deviation of the {\em local proton\/} spectrum
from this slope is naturally explained by the influence of a local, recent
source~\cite{LS}. Therefore we use a universal $1/E^{2.2}$ power-law 
as injection spectrum for all nuclei. Inside galaxies, 
the spectrum is modified because of energy-dependent CR confinement 
in them, whereas the CR flux exiting them retains the original injection 
$1/E^{2.2}$ spectrum. 
We assume that spiral galaxies have magnetic fields 
which are on average similar to the one in the Milky Way. 
Hence we can use for the confinement time $\tau(E)\propto X(E)$ of CRs
in normal spiral galaxies the grammage $X(E)$ calculated in Refs.~\cite{PI,PII}
for the Milky Way.

In contrast, the observed strengths of magnetic fields in starburst 
galaxies are a factor $\sim 100$ larger than in the Milky 
Way~\cite{Lacki:2013ry}. Since the confinement time $\tau(E)$ is a function of $E/(ZeB)$, we can compensate for this change in $B$ by 
rescaling the CR energy $E$ in the grammage calculated 
for the Milky Way. We assume that the coherence length $l_{\rm c}$ 
of the turbulent interstellar magnetic fields inside starburst galaxies is not 
significantly different from that in normal galaxies.

\begin{figure}
  \includegraphics[width=0.45\textwidth,angle=0]{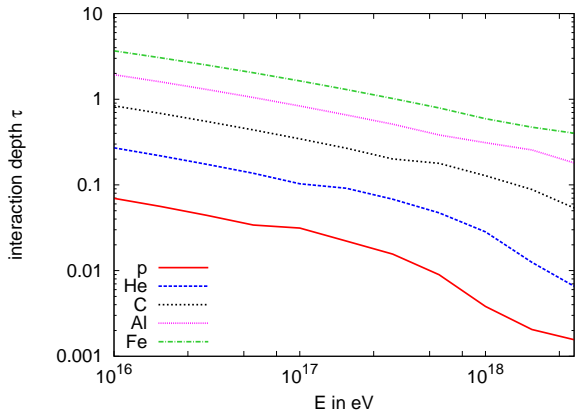}
  \caption{Interaction depth $\tau$ in a starburst galaxy, as a function of CR energy $E$, 
           for Fe (top, green line), Al (magenta), C (black), He (blue) and protons
           (bottom, red). 
     \label{figtau}}
\end{figure}

In Fig.~\ref{figtau}, we show the resulting interaction depth 
$\tau=\sigma_{\rm inel} X /m_p$ in a starburst galaxy for various CR 
nuclei as a function of energy, where we used the density
profile described in~\cite{PI} with $n_0=1/$cm$^3$. 
The fraction of CRs which are absorbed is given by 
$f_{\rm int} =1-\exp(-\tau)$. Thus the flux of heavy 
nuclei, such as Al or Fe, leaving a starburst galaxy is exponentially 
suppressed. In contrast, CR interactions on gas can be neglected
in normal galaxies because of their weaker magnetic fields and shorter 
CR confinement times.

Let us assume that every normal spiral galaxy (resp.\ every starburst 
galaxy) accelerates CRs with charge $Z$ up to a maximum 
energy $E_{\max}=Z \times 10^{17}$\,eV (resp.\ $E_{\max}=Z \times 10^{18}$\,eV). 
The larger maximum energy achievable in starburst galaxies may e.g.\ be
connected to the larger magnetic fields, which makes an additional
acceleration of CRs in superbubbles more efficient~\cite{superbubbles}.
Finally, we assume that the fractions of the injected CR nuclei are the 
same as the ones below the knee in the Milky Way, except for 
the proton fraction. We assume that the local proton 
fraction measured just below the knee is reduced due to the steeper proton spectrum compared to the 
average one in normal galaxies. We choose to set the proton fraction to $f_p=0.5$ of 
the CR flux. Thus the composition of p:He:N:Al:Fe is fixed as 
50:22.5:12.5:5:10, which is consistent with the one deduced in~\cite{PII} for
He:N:Al:Fe at $E=10^{14}$\,eV.

\subsection{Diffuse fluxes from normal spiral and 
starburst galaxies}
\label{CR}

\begin{figure}
\includegraphics[width=\columnwidth]{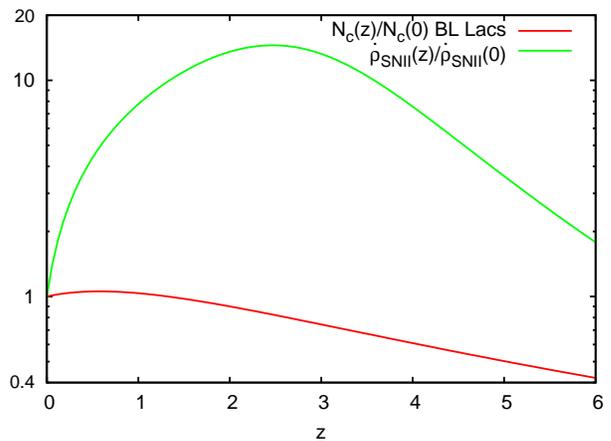}
\caption{Rate $\dot\rho$ of type II SNe (green curve) and effective comoving source 
density $N_c(z)$ of BL Lacs (red) as a function of redshift, normalized to their 
present values at $z=0$.}
\label{fig:evolution}
\end{figure}

We compute now the diffuse flux of extragalactic CRs from all star-forming 
galaxies.
Let us first assume that at redshift $z$ the CR emissivity $Q_{\rm CR}(z)$ scales with the Type II supernova (SN) rate $\dot{\rho}_{\rm SNII}(z)$, and thus with the star-formation rate (SFR). This is a phenomenological argument, which does not directly imply that these high-energy CRs have been accelerated at supernova shock waves --see~\cite{PII} for a discussion on possible acceleration mechanisms. For $\dot{\rho}_{\rm SNII}(z)$, we use the parametrization presented in~\cite{Hopkins:2006bw} and take the Baldry \& Glazebrook initial mass function~\cite{Baldry:2003xi}. The corresponding Type II SN rate is~\cite{Hopkins:2006bw} 
\be
 \dot{\rho}_{\rm SNII}(z) = \frac{0.0132 \, (a+bz) \, h}{1+(z/c)^{d}} \; {\rm yr}^{-1}{\rm Mpc}^{-3} \,,
\ee
with $a=0.0118$, $b=0.08$, $c=3.3$, $d=5.2$ and $h=0.7$. It is plotted in 
Fig.~\ref{fig:evolution} as a green line, which peaks at redshift $z\sim 2-3$. 
We assume that, globally, a fraction $\epsilon_{\rm CR} = 0.1$ of the kinetic energy of supernovae ($E_{\rm SN} \approx 10^{51}$\,erg per SN) is channeled into CRs. The integral CR emissivity $Q_{\rm CR}(z)$ is then
\be
 Q_{\rm CR}(z) \simeq 9 \times 10^{-22} \, \frac{(a+bz) \, h}{1+(z/c)^{d}} \; \frac{{\rm eV}}{{\rm cm^{3}\,s}} \,.
\label{QCR}
\ee

We define $Q_{\rm CR,SB}$ (resp. $Q_{\rm CR,SP}$) as the integral CR emissivity due to starburst (resp. normal spiral) galaxies. $Q_{\rm CR,SB} = f_{\rm SB}\,Q_{\rm CR}$ and $Q_{\rm CR,SP} = (1-f_{\rm SB})\,Q_{\rm CR}$, where $f_{\rm SB}$ denotes the fraction of SFR (or SNe) occurring in starburst galaxies at redshift $z$. 
We parametrize $f_{\rm SB}$ with the two examples proposed in~\cite{Thompson:2006np} : $(i)$ $f_{\rm SB} = 0.9\,z + 0.1$ at $z \leq 1$, and $=1$ otherwise, or $(ii)$ $f_{\rm SB} = 0.1\,(1+z)^{3}$ at $z \leq 1$, and $=0.8$ otherwise. We also consider a scenario $(iii)$ motivated by Ref.~\cite{Gruppioni:2013jna}, where star-forming galaxies are divided into four classes: normal spiral galaxies, starburst galaxies and star-forming AGNs. The latter category is divided into two subsets: SF-AGNs which resemble spiral galaxies and SF-AGNs which resemble starburst galaxies.

We can now determine the differential CR emissivity $q_{\rm CR,SB}(E,z)$ from starburst galaxies, with
\be
Q_{\rm CR,SB}(z) = \int dE \,E q_{\rm CR,SB}(E,z) = \int dE \,E \dot n(z) \, \frac{dN}{dE} \, ,
\ee
taking an injection spectrum $\frac{dN}{dE}\propto E^{-2.2}$, between $E_{\min}=1$\,GeV and $E_{\max}=10^{18}$\,eV. This yields 
\be \label{q}
q_{\rm CR,i}(E,z) \simeq \frac{1 \times 10^{-20}}{{\rm eV\,cm^{3}\,s}} \, f_{i} \, \frac{(a+bz) \, h}{1+(z/c)^{d}} \left( \frac{E}{1\,{\rm eV}} \right)^{-2.2} \, ,
\ee
with  $i=\{{\rm SB},{\rm SP}\}$.

We can find the resulting diffuse CR intensity $I(E)$ from 
\be \label{Idiff}
 I(E)= \frac{c}{4\pi H_0} \int_0^{z_{\max}}
        \frac{dz}{(1+z)\omega} \, q_{\rm CR}(z,E') \, \e^{-\tau(E')} \, ,
\ee
where $\omega=\sqrt{\Omega_\Lambda+\Omega_{\rm m}(1+z)^3}$. In the following, we take $H_0=70$\,km\,s$^{-1}$\,Mpc$^{-1}$, $\Omega_{\Lambda}=0.7$ and $\Omega_{\rm m}=0.3$. Below $10^{18}$\,eV, we can neglect CR interactions in the intergalactic space and thus set $E'= (1+z)E$. For such CR energies, we can neglect absorption during propagation, and $\tau$ corresponds to the one  given by Fig.~\ref{figtau}. The upper integration limit $z_{\max}$ is given by the magnetic horizon. 
In the following, we consider the optimistic case where the intergalactic magnetic fields are sufficiently weak to have $z_{\max} \geq 6$ for $E \geq 10^{16}$\,eV. 
This case of no magnetic horizon yields an upper limit on the diffuse extragalactic CR flux one can expect from all star-forming galaxies. 
We set $\dot{\rho}_{\rm SNII}(z)$ to zero at redshifts $z > 6$, as in~\cite{Hopkins:2006bw}. Then,
\bd
 I_{\rm SB}(E)= \frac{c}{4\pi H_0} \times \frac{10^{-20}}{{\rm eV\,cm^{3}\,s}} \, \left( \frac{E}{1\,{\rm eV}} \right)^{-2.2}
\ed
\bd
 \times \int_0^{z_{\max}}\frac{dz}{\omega} \, f_{\rm SB} \, (1+z)^{-3.2} \, \frac{(a+bz) \, h}{1+(z/c)^{d}} \, .
\ed

One finds for the total diffuse CR intensity $I(E)$,
\be
 E^{2} \, I(E) \simeq \frac{1 \times 10^{2} \, {\rm eV}}{{\rm cm^{2}\,s\;sr}} \left( \frac{E}{10^{15}\,{\rm eV}} \right)^{-0.2} \, ,
\ee
and $I_{\rm SB}(E) \simeq 0.63 \,I(E)$ (resp. $I_{\rm SB}(E) \simeq 0.46 \,I(E)$) in scenario $(i)$ (resp. scenario $(ii)$). With $f_p=0.5$, this gives a CR proton flux that is more than one order of magnitude weaker than the extragalactic CR proton flux deduced for the escape model: See the orange line in Figure 11 of Ref.~\cite{PII}. This shows that the guaranteed contribution from all star-forming galaxies in the escape model to the diffuse CR flux is negligible.
Note that this conclusion does not depend on our assumption about the maximal
acceleration energy in star-forming and starburst galaxies.

\begin{figure*}
  \centerline{\includegraphics[width=0.49\textwidth,angle=0]{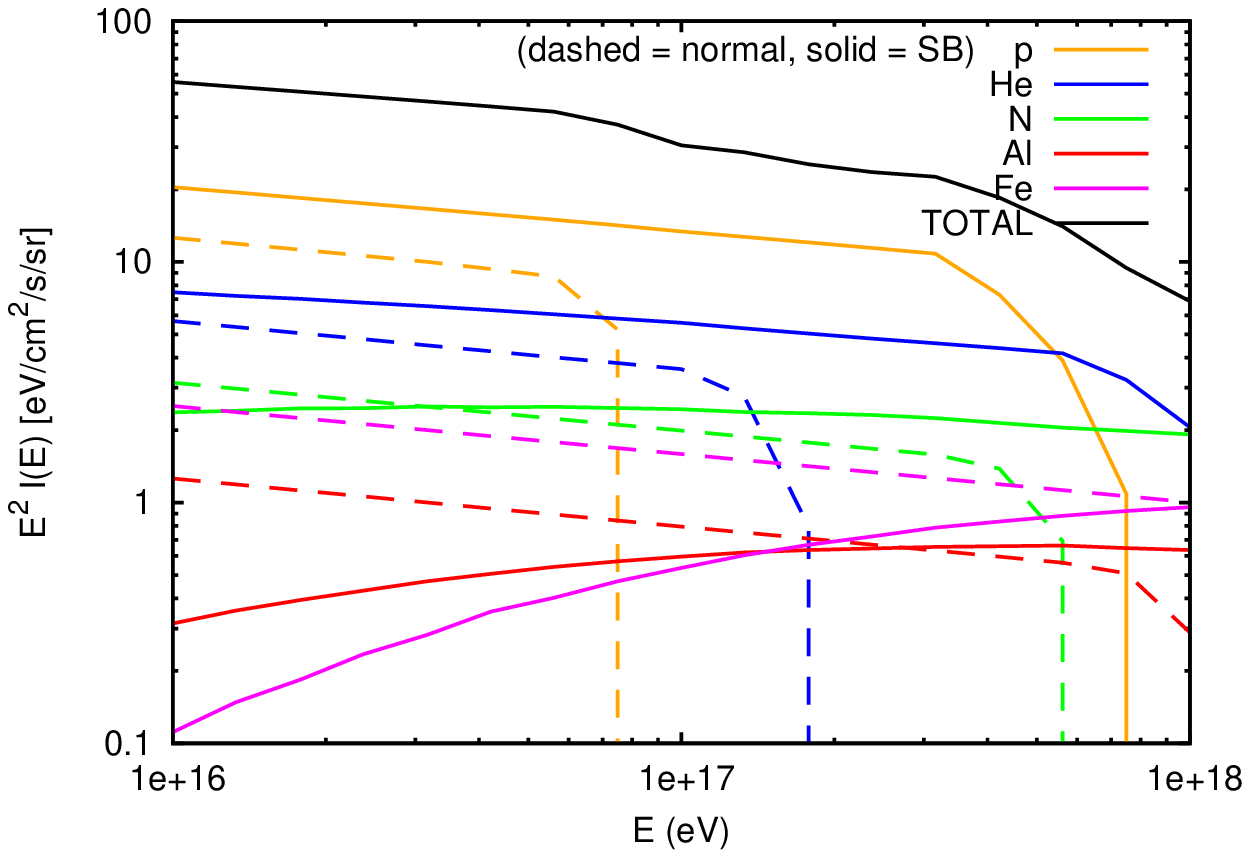}
              \hfil
              \includegraphics[width=0.49\textwidth,angle=0]{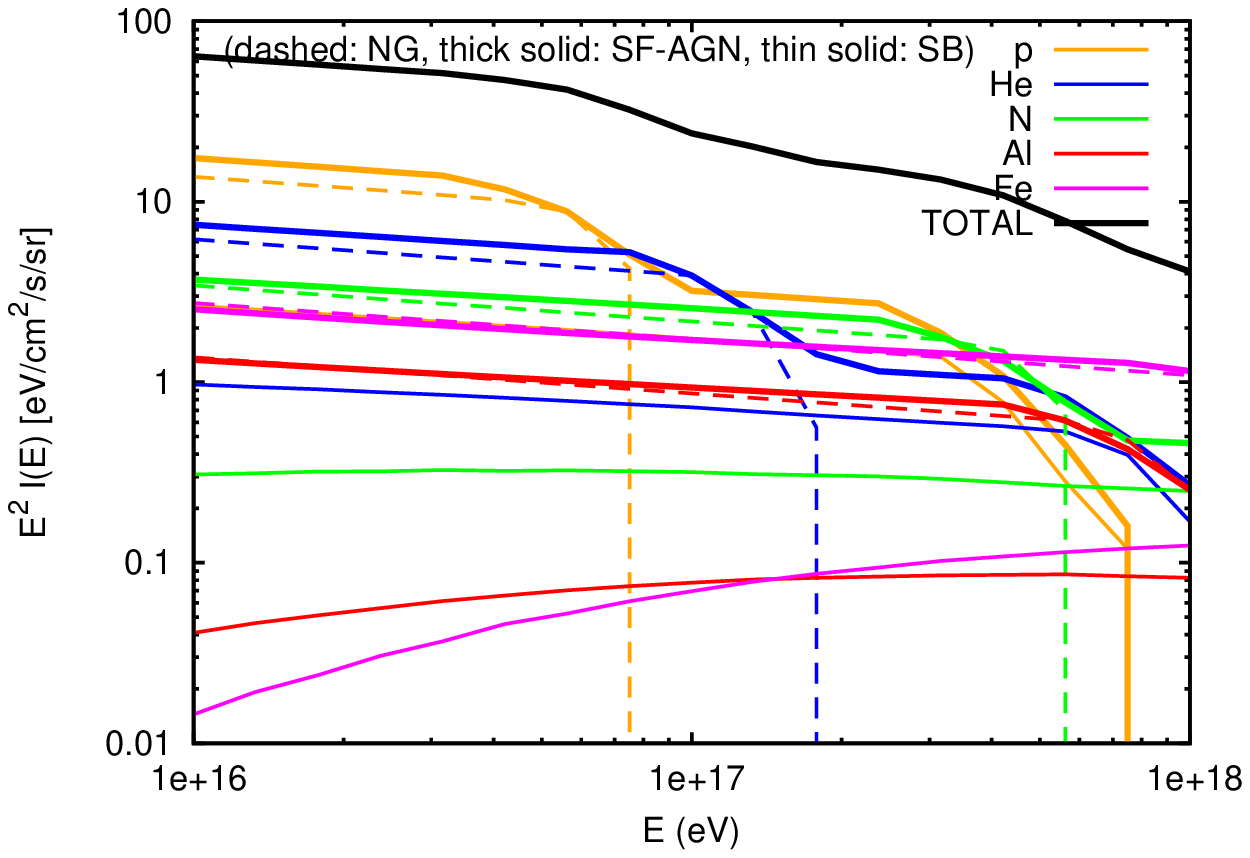}
             }
  \caption{{\it Left panel:} Diffuse CR flux from normal spiral (dashed lines) and starburst galaxies
 (solid lines) within scenario (i), as a function of energy. Total flux in black. Individual contributions of 
protons (orange), He (blue), N (green), Al (red) and Fe (magenta). {\it Right panel:} Diffuse CR flux from normal spirals (dashed lines), star-forming AGNs --both starburst and non-starburst ones-- (thick solid line) and starburst galaxies (thin solid lines) as a function of energy, within scenario (iii). For both panels, IGMF with strength $B=10^{-17}$\,G and coherence length $l_c=1$\,Mpc. 
\label{figDiffuseCR}}
\end{figure*}

Using the above parameters, we present in Fig.~\ref{figDiffuseCR} (left panel) the diffuse fluxes of CR nuclei due to normal and starburst galaxies between $E=10^{16}$\,eV and $E=10^{18}$\,eV. In this computation, we use scenario $(i)$ for the evolution with redshift of the fraction of SNe occurring in starburst galaxies. We consider IGMFs with strength $B=10^{-17}$\,G and coherence length $l_c=1$\,Mpc,  e.g.\ values which are consistent with lower limits from \gr\ observations for time-varying sources~\cite{B17}. With such parameters, there is no magnetic horizon for any of the nuclei at the energies we consider. The individual contributions of protons and nuclei from normal spiral (resp. starburst) galaxies are shown with the dashed (resp. solid) lines below the total flux. Orange lines for protons, blue ones for helium, green ones for CNO nuclei, red ones for aluminium and magenta ones for iron. The fluxes of intermediate and heavy nuclei from starburst galaxies are exponentially suppressed at the lowest energies, because of nuclei suffering significant energy losses on background gas in starbursts, see Fig.~\ref{figtau}. In Fig.~\ref{figDiffuseCR} (right panel), we show, for comparison, the results for scenario $(iii)$, where the CR flux is dominated by SF-AGNs.

\begin{figure}
  \centerline{
  \includegraphics[width=0.49\textwidth,angle=0]{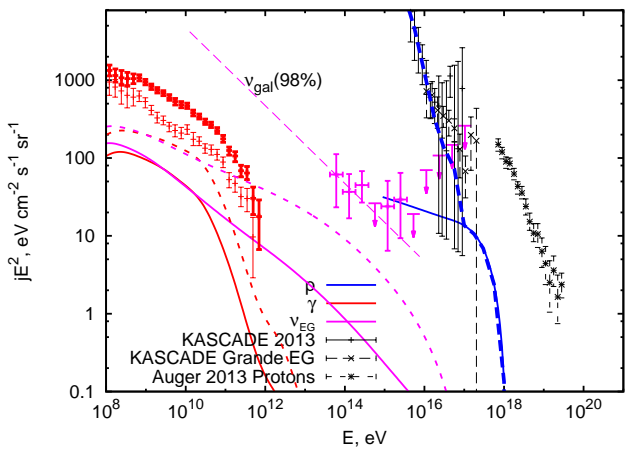}
             }
  \caption{Diffuse CR proton (blue), gamma-ray (red) and neutrino (magenta) fluxes from starburst galaxies together with CR proton data from KASCADE, 
KASCADE-Grande~\cite{dataKG} and Auger (black errorbars)~\cite{Aab:2013ika,Aab:2014aea}, the IGRB and the EGB from Fermi-LAT (red errorbars)~\cite{Fermi}, and high-energy neutrino flux from IceCube (magenta errorbars)~\cite{Ice}. Solid red and magenta lines for gamma-ray and neutrino fluxes with $n=1$\,cm$^{-3}$, and dashed lines for $n=10$\,cm$^{-3}$.
\label{FluxesStarbursts}}
\end{figure}

We consider next the contribution from star-forming galaxies to the 
diffuse neutrino and \gr\ fluxes. 
The fraction of CR protons that interact in starburst galaxies is given by 
$f_{\rm int} =1-\exp(-\tau)$, where $\tau$ is the interaction depth for
protons shown in Fig.~\ref{figtau}. 
We present in Fig.~\ref{FluxesStarbursts} the resulting gamma-ray (red line) 
and neutrino (magenta) flux from starburst galaxies, within evolution
scenario $(i)$. 
We also plot the diffuse CR proton flux from starburst galaxies. 
First, we note that star-forming galaxies give a subdominant 
contribution to the primary CR flux, for the spectral index $\alpha_p = 2.2$ 
which is favored by the escape model. This discrepancy could be reduced 
by increasing e.g.\ the fraction $\eps_{\rm CR}$ of energy transferred to 
CRs or the SN rate. However, the redshift evolution of star-forming 
galaxies leads to a proton flux which spectral shape disagrees with 
the shape deduced for the extragalactic proton flux. Thus we conclude 
that star-forming galaxies cannot be the main contributor to the 
extragalactic proton flux (up to their $E_{\max}$).
Next, we compute the secondary neutrino and photon fluxes. We 
compare them respectively to the astrophysical neutrino flux (magenta 
errorbars) measured by IceCube~\cite{Ice} and to the measurement (red 
errorbars) of the extragalactic  \gr\ background (EGB) by 
Fermi-LAT~\cite{Fermi}. For the latter, we show both 
the IGRB (lower curve) and the total EGB including resolved sources (upper 
curve). Choosing  as average gas density $n=1/$cm$^3$, star-forming 
galaxies contribute around 30\% at 10\,GeV to the IGRB, while their 
contribution to the neutrino  signal observed by IceCube reaches 10\% below
 $10^{14}$\,eV. Increasing the gas density by a factor of ten leads only to 
an  increase of the secondary fluxes by a factor a few, because the source
is already thick in a large energy range.
Taking into account the uncertainty in the grammage used, 
we conclude that star-forming galaxies cannot explain the extragalactic 
proton flux, but may
contribute a significant fraction to the IGRB and to the extragalactic 
part of the neutrino signal as observed by IceCube, especially at 
low energies.

Not surprisingly, studies which assume $\alpha_p \simeq 2.0$ and a large 
grammage in starburst galaxies can reproduce a high-energy neutrino 
flux comparable to that of IceCube, 
see e.g.~\cite{Liu:2013wia,Tamborra:2014xia}.

\section{UHECR sources}

We discuss next AGNs, which are generally considered to be prime candidates 
for the sources of UHECRs. We consider the subset of 
radio-loud AGN or, more precisely, the BL Lac/Fanaroff-Riley I (FR I) 
sub-class of the radio-loud AGN population. This choice is motivated 
by two reasons: First, the evolution of these sources is relatively slow 
and peaks at low redshift. Thus, the resulting diffuse CR flux has a 
rather different spectral shape than the one of star-forming galaxies. 
Second, BL Lacs have been suggested to be a major contributor to the IGRB (see e.g.~\cite{Neronov:2011kg,dimauro,Ajello:2015mfa}), which raises the question 
whether these sources can also fit at the same time the extragalactic CR proton flux expected in the escape model. 
Also, AGNs are natural candidates of high-energy neutrino sources, see amongst others 
Refs.~\cite{Stecker:1991vm,Kalashev:2014vya}, and Ref.~\cite{Murase:2014foa} 
for neutrino production in the inner jets of radio-loud AGNs, 
including blazars, as sources of UHECRs.

\subsection{Evolution of BL Lacs}

We determine the cosmological evolution of BL Lac/FR I sources from the
corresponding evolution of the $\gamma$-ray luminosity, assuming that the CR
and the \gr\ luminosity are proportional,
\begin{equation}
N_c(z) \propto \int_{L_{\gamma}^{min}}^{L_{\gamma}^{max}} \rho(z,L_{\gamma}) L_{\gamma} dL_{\gamma} .
\end{equation}
Here, $\rho(z,L_{\gamma})$ is the $\gamma$-ray luminosity function (LF), i.e.\ the
number of sources per  comoving volume and luminosity. For $\rho(z,L_{\gamma})$ 
we adopt the Luminosity-Dependent Density Evolution (LDDE) model of 
Ref.~\cite{dimauro}. Within this model, the LF $\rho(z, L_{\gamma})$ can be 
expressed as
\begin{eqnarray}
\label{eq:ldde1}
   \rho(z,L_{\gamma})= \rho(L_{\gamma}) \, e(z,L_{\gamma}) ,
\end{eqnarray}
with 
\begin{eqnarray}
\label{eq:ldde0}
   \rho(L_{\gamma})= 
	\frac{A}{\log{(10)}L_{\gamma}}\left[ \left( \frac{L_{\gamma}}{L_{c}} \right)^{\gamma_1} + \left( \frac{L_{\gamma}}{L_{c}} \right)^{\gamma_2} \right]^{-1} ,
\label{eq:ldde2}\\
   e(z,L_{\gamma}) = \left[ \left( \frac{1+z}{1+z_c(L_{\gamma})} \right)^{p_1} + \left( \frac{1+z}{1+z_c(L_{\gamma})} \right)^{p_2} \right]^{-1} ,
\end{eqnarray}
and
\begin{eqnarray}
\label{eq:ldde3}
  z_c(L_{\gamma}) = z^{\star}_c\, \left( \frac{L_{\gamma}}{10^{48} \rm{\,erg\,s}^{-1}}\right)^{\alpha} \,.
\end{eqnarray}
The numerical values for the parameters were determined in~\cite{dimauro} 
from a fit to the statistics of BL Lacs observed by the Fermi-LAT telescope, cf.\ their Table 3. 
The evolution of the effective source density with the redshift is shown in 
Fig.~\ref{fig:evolution}. 
In contrast to average AGNs, the number density of BL Lac and FR I galaxies 
peaks at low redshift, $z\lesssim 1$. Their evolution is similar to that 
of  galaxy clusters. In fact, most of the FR I sources, which are in the unified AGN scheme the same 
sources as BL Lacs seen under different observation angles, reside in the centres of the dominant central 
elliptical galaxies of galaxy clusters (cD galaxies).

\subsection{Interactions in BL Lac/FR I sources}

We assume that the CR injection spectrum of each source follows 
a power-law with slope $\alpha_p$ and exponential cut-off,
\begin{equation}
 \frac{dN_{CR}}{dE}\propto 
 E^{-\alpha_p}\exp\left(-\frac{E}{E_{\rm cut}}\right) \,.
\end{equation}
For each assumed slope  $\alpha_p$ of the spectrum, we adjust the 
cutoff energy $E_{\rm cut}$ in such a way that the spectrum of the entire 
source population (integrated over redshift) fits best the observed cosmic ray 
spectrum in the energy range $10^{17}$\,eV -- $10^{20}$\,eV.

Cosmic rays of low energy are not necessarily escaping from the source. 
First of all, they could be trapped right in the source. 
The condition of free escape from the 
source is that the Larmor radius of the accelerated particle is comparable to 
the source size $R$. This condition reads $E\gtrsim E_{\rm free}$, where 
\begin{equation} 
 E_{\rm free}\simeq  eBR
 \simeq 3\times 10^{20}\,{\rm eV}\;
 \frac{B}{10^4\mbox{ G}}\;\frac{R}{10^{14}\mbox{ cm}} \,,
\end{equation}
$e$ is electric charge of the particle and $B$ is the  magnetic field 
strength. Lower energy particles are trapped inside the source, be it the 
AGN central engine, jet or the radio lobes. 

The trapped particles can still escape from the source, but 
in a diffusive way on a much longer time scale. The details of this process 
depend on the turbulence of the magnetic field in the relevant source 
structure. The time scale of turbulence development on a distance scale 
$\lambda$ can be estimated by the 
eddy turnover time $T_{\rm turb}\sim \lambda/v$, where $v$ is the average bulk 
velocity of the plasma moving over the distances of the order of $\lambda$. 
In the case of the central engine of AGN, this velocity scale is the typical 
velocity  of the accretion flow, which is about the Keplerian or free-fall 
velocity. Close to the black hole horizon, this velocity is relativistic, 
$v\sim c$.
In the AGN jet, the velocity is also $v\sim c$  because the jet is a 
relativistic outflow. Only in the case of the large scale radio lobes the 
velocity could be $v\ll c$. In this case it is determined by the details of 
interaction of the lobes with the interstellar/intracluster medium. In any 
case, the eddy turnover scale is certainly much shorter than the source 
lifetime for all the elements of the radio loud AGN. This means that the 
medium and magnetic field in the source are turbulent.

The turbulence power spectrum may for example follow a Kolmogorov 
or an Iroshnikov-Kraichnan power law. In our calculations, we assume that the 
power spectrum of the turbulence is a Kolmogorov one. 
Then the escape time of CRs scales with energy as 
\begin{equation}
\label{eq:tesc}
 t_{\rm esc} = \frac{R}{c}\left(\frac{E}{E_{\rm free}}\right)^{-1/3}
 \simeq 5\times 10^{6} \mbox{ s}\left[\frac{E}{10^{11}\mbox{ eV}}\right]^{-1/3} \,,
\end{equation}
where we used as source size $R=10^{14}$\,cm and as magnetic field strength $B=10^4$\,G.

Cosmic rays trapped inside the source lose energy by interacting with the ambient medium present in the source. In the case of the accretion flow, the density is moderately low $n\lesssim 10^{10}$~cm$^{-3}$ for the radiatively inefficient accretion flows powering FR I/BL Lac sources. The energy loss time of CR protons is
\begin{equation}
t_{pp}= \frac{1}{c\kappa\sigma_{pp}n}\simeq 1\times 10^6\mbox{ s}
 \;\left( \frac{n}{10^{9}\mbox{ cm}^{-3}} \right)^{-1} \,,
\end{equation}
where $\sigma_{pp}\sim (3-8)\times 10^{-26}$\,cm$^2$  is the inelastic $pp$ cross-section and $\kappa\simeq 0.6$  the inelasticity. The interaction time is shorter than the escape time, $t_{\rm esc}\gtrsim t_{pp}$,  for CRs with  energy $E<E_{\rm esc}=8\times 10^{12}$\,eV, where we used again $R=10^{14}$\,cm and $B=10^4$\,G for
the numerical estimate. Thus CRs with energies below $\sim 10$\,TeV would 
not escape from the central engine of an AGN powered by a 
$3\times 10^8M_\odot$ black hole. 
Note however that the numerical value of the escape energy depends strongly
on the chosen values for $n$, $B$ and $R$ and should be considered
therefore only as an indication.

\begin{figure}
\includegraphics[width=\columnwidth]{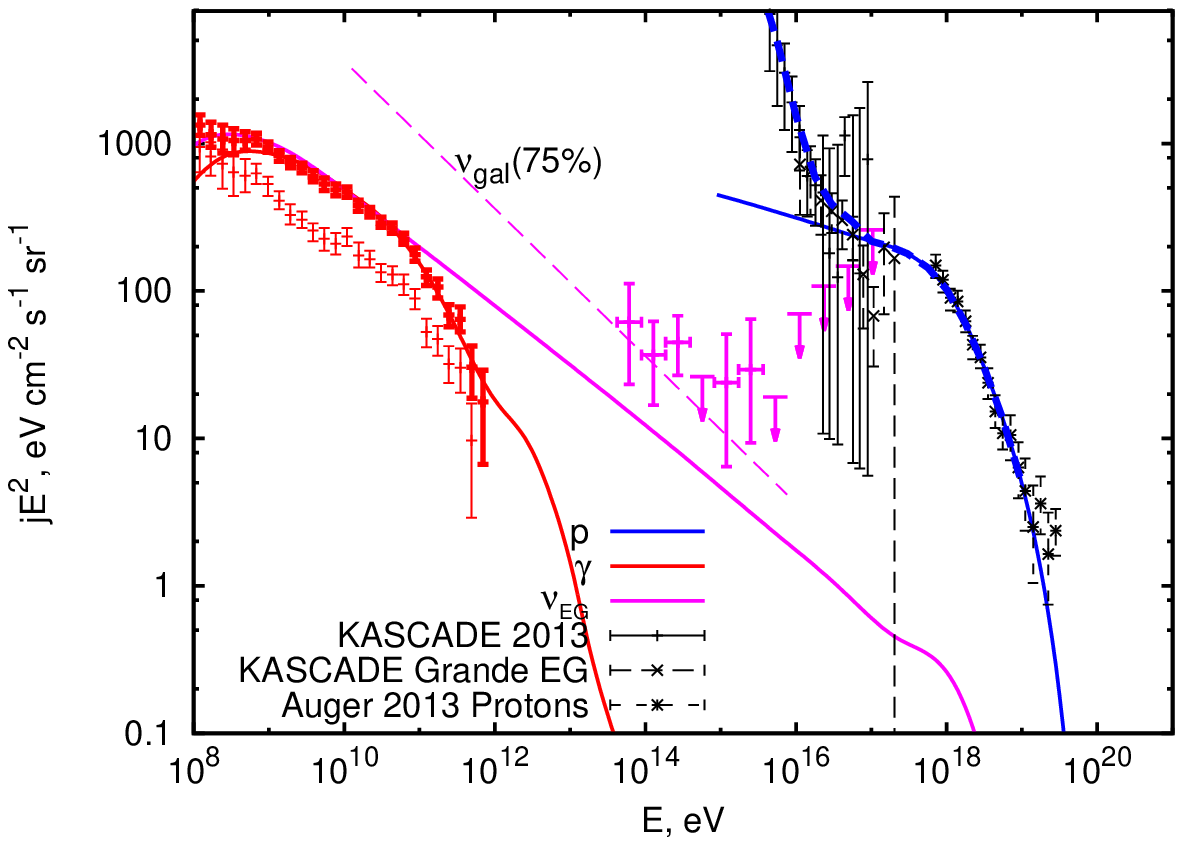}
\includegraphics[width=\columnwidth]{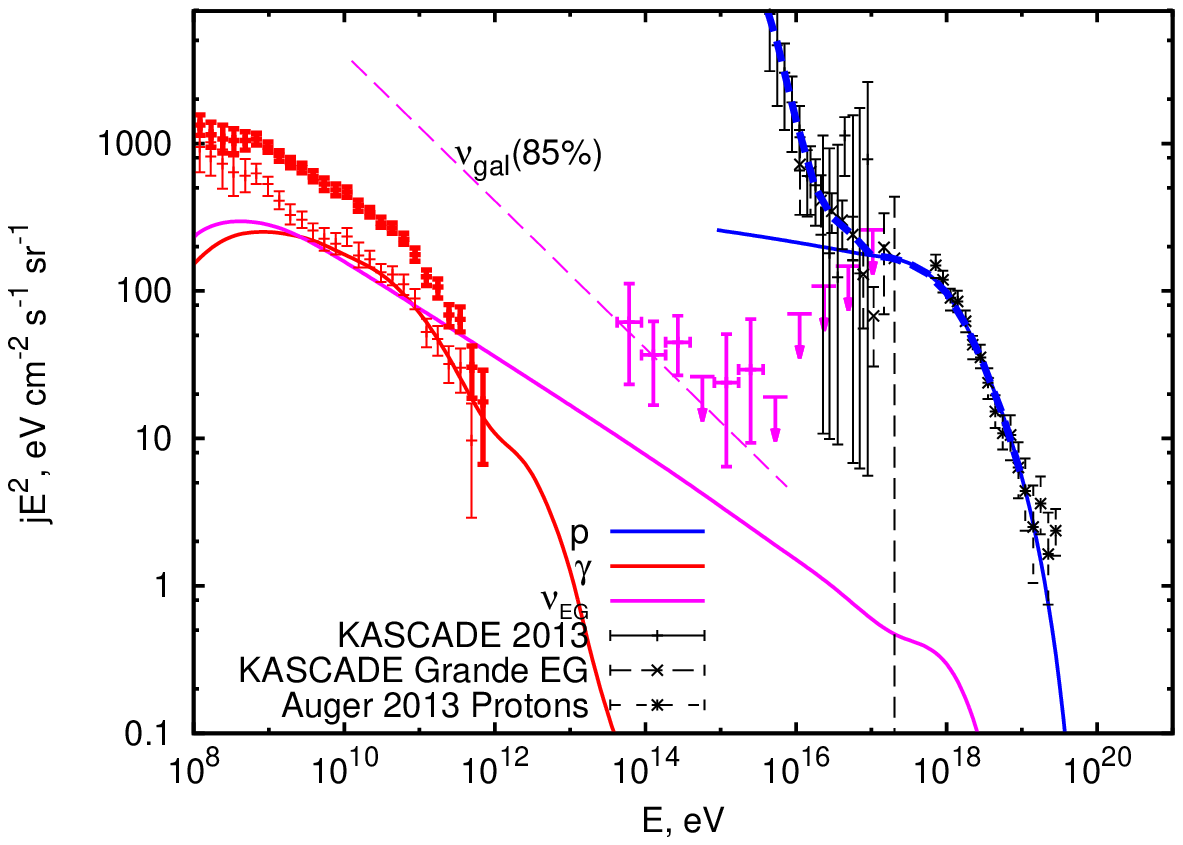}
\includegraphics[width=\columnwidth]{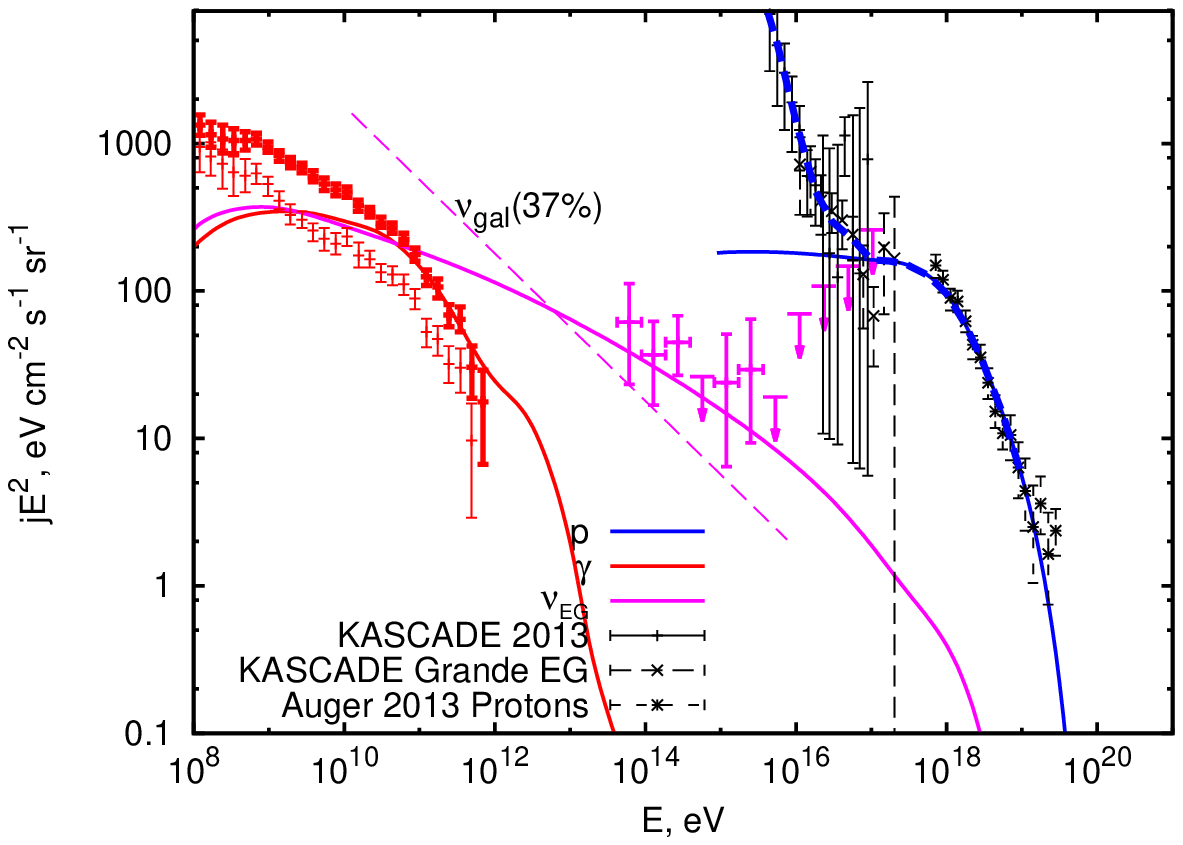}
\caption{Diffuse flux of CR protons from  BL Lacs (thick blue line),
Galactic proton flux in the escape model (thin blue line) together with
the resulting photon (red line) and neutrino (magenta) fluxes. Upper panel: $\alpha_p=2.17$ and $E_{\rm esc}=3 \times 10^{11}$\,eV; 
Middle panel: $\alpha_p=2.10$ and $E_{\rm esc}=3 \times 10^{11}$\,eV; Lower panel: $\alpha_p=2.10$ and $E_{\rm esc}= 10^{14}$\,eV. 
$E_{\max}=10^{19}$\,eV for all three panels. CR proton data from KASCADE, 
KASCADE-Grande~\cite{dataKG} and Auger (black errorbars)~\cite{Aab:2013ika,Aab:2014aea}. IGRB and EGB from Fermi-LAT (red errorbars)~\cite{Fermi} 
and high-energy neutrino flux from IceCube (magenta errorbars)~\cite{Ice}.}
\label{fig:spectrum}
\end{figure}

Our results for the diffuse flux of CR protons from  UHECR sources following 
the BL Lac evolution (\ref{eq:ldde0}) is shown in Fig.~\ref{fig:spectrum}
for $E_{\rm cut}=10^{19}$\,eV, and two different values of CR slope, 
$\alpha_p=2.17$ 
for the upper panel and $\alpha_p=2.10$ for the middle and lower panels. 
The Galactic proton flux in the escape model is shown with a dashed blue line. 
The choice $\alpha_p=2.17$ (resp. $\alpha_p=2.1$) results in an excellent (resp. good)
fit of the extragalactic proton component deduced from Auger and 
KASCADE-Grande measurements. 

In the same figure, we also show the secondary 
fluxes obtained for $E_{\rm esc}=3 \times 10^{11}$\,eV (upper and middle panels) and 
$E_{\rm esc}=10^{14}$\,eV (lower panel), 
i.e.\ values of $E_{\rm esc}$ which are characteristic for
CR acceleration close to the supermassive BH powering the BL Lac.
The diffuse photon (resp. neutrino) fluxes are shown 
with red (resp. magenta) lines.

One can see in the upper panel that for $\alpha_p=2.17$ and 
$E_{\rm esc}=3 \times 10^{11}$\,eV, the photon flux from the BL Lac/FR I 
populations 
may explain the entire extragalactic \gr\ background. 
This choice of parameters would imply that the main part of the observed 
TeV \gr\ is of hadronic origin. The synchrotron peak observed in the
spectra of BL Lacs at lower energies is caused in this picture by electrons
which can escape from the central engine and radiate most of their energy
in the weaker magnetic field of the surrounding host galaxy.
Note that the agreement of the observed
and the predicted \gr\ {\bf flux} is non-trivial, because the model parameters 
were chosen to fit the UHECR protons, rather than the IGRB spectrum. The predicted 
high-energy neutrino flux from these AGNs is about a quarter of the IceCube neutrino flux, 
requiring a Galactic contribution to these neutrinos at the level of 75\% (dashed magenta line). 
The diffuse \gr\ and neutrino fluxes at Earth are due to in-situ 
production and cascade emission during CR propagation. We show in the figure both the 
EGB and the IGRB. One should note that as long as the in-situ emission 
dominates, \gr s (and neutrinos) point back to their sources 
and the \gr\ emission is then truely part of the EGB if the source is detected, 
and not part of the IGRB. If extragalactic 
magnetic fields have a negligible impact, the distinction would become irrelevant.

For $\alpha_p=2.1$ and the same grammage (middle panel), 
the photon and neutrino fluxes are somewhat lower. Indeed, for a CR harder spectrum and 
for an extragalactic CR flux that satisfies the 
levels observed at very high energies, less low energy CRs are present. 
For $\alpha_p=2.1$ and $E_{\rm esc}=3 \times 10^{11}$\,eV, the photon flux from the BL Lac/FR I populations 
provides a good fit to the IGRB deduced by Fermi-LAT~\cite{Fermi}, as can be seen in the middle panel.

The impact of $E_{\rm esc}$ on the secondary fluxes can be seen comparing
the middle to the lower panel of Fig.~\ref{fig:spectrum}: 
For the same slope $\alpha_p=2.1$,
a change in the value of $E_{\rm esc}$ affects the resulting diffuse \gr\ flux 
much weaker than the high-energy neutrino flux. The latter are produced
by CRs whose interaction depth is $\tau\ll 1$. In this regime, the secondary
fluxes scale linearly with the grammage. In contrast, the contribution
to the diffuse \gr\ flux of CRs with $\tau\gg 1$ is practically not 
affected by a change in  $E_{\rm esc}$.  For the parameters chosen
in the lower panel, one can now explain about $\sim 60\%$ of the IceCube 
flux by secondary neutrinos from BL Lacs/FR Is.

Explaining the extragalactic CR flux within the escape model, the IGRB/EGB 
and a large fraction of IceCube neutrinos requires a sufficiently large 
interaction depth at the source (as that chosen in the lower panel). 
Therefore our scenario favors CR acceleration close to the black hole 
of the BL Lacs/FR Is,  in a region where $pp$ interactions dominate 
over $p\gamma$ interactions, such as the polar caps. Let us however note that other sites of CR 
acceleration such as the jets or lobes of BL Lacs/FR Is can still explain 
the IGRB/EGB flux, provided that $\alpha_p$ is somewhat larger than 2.2. 
In this case, the contribution of these sources to the IceCube neutrino flux 
would be however reduced.

\subsection{Neutrinos from BL Lacs and IceCube neutrinos}

A remarkable feature of the secondary spectra is that the neutrino flux 
for our reference parameters can be comparable to the flux of astrophysical 
neutrinos measured by IceCube~\cite{Ice}, see especially Fig.~\ref{fig:spectrum} (lower panel).
In Fig.~\ref{fig:spectrum}, the dashed magenta lines indicate 
the Galactic neutrino flux required\footnote{We estimate the required 
Galactic contribution assuming an 
$E^{-2.5}$ spectrum of the Galactic neutrino flux and normalizing the
total neutrino flux using the first energy bin of the IceCube data at
$6\times10^{13}$\,eV.} to match 
the observed IceCube signal: The additional Galactic neutrino contribution 
varies between $\simeq 90$\% and $\simeq 30$\%, depending on the value 
of $\alpha_p$. This is in line with the results from Refs.~\cite{Neronov:2013lza}, which 
found a high-energy neutrino flux from our Galaxy at the level of $\sim 50$\% of the IceCube flux, 
taking a global Galactic cosmic ray spectrum slope of 2.5 instead of 2.7 (local flux).

Also, the slope of our neutrino spectrum is close to the one measured by 
IceCube. The slope of the extragalactic neutrino spectrum in the IceCube 
range is $\alpha_\nu\simeq \alpha_p+\delta\simeq 2.4 - 2.5$ for an injection 
spectrum with $\alpha_p\simeq 2.1 - 2.2$ and Kolmogorov turbulence, 
$\delta=1/3$. This is an important difference with respect to other  
models that predict a harder spectrum, with slopes $\approx 2.0$, i.e.\ 
similar to the slope of the injection spectrum of CRs at the sources. In 
such models, either the parent CRs escape freely from the sources or lose 
all their energy in the sources. In our model, CRs  diffuse in the source 
before escaping, which results in an additional softening by $\delta$ of 
the neutrino slope. 

Therefore, our model can explain the entire astrophysical neutrino signal 
observed by IceCube, both in terms of the flux level (Galactic and 
extragalactic contributions of the same order of magnitude) and of the
slope.

The increasing size of the neutrino sample with time will allow
IceCube to constrain the ratio of the Galactic and extragalactic high-energy 
neutrino fluxes, studying the anisotropy of their arrival directions. 
Within our model, this flux ratio depends strongly  on the
slope of the extragalactic proton flux and provides therefore an
important information on the extragalactic UHECR sources.

\subsection{Interactions in the host galaxy and galaxy cluster}

We verify in this subsection which impact the host galaxy and galaxy cluster have 
on the diffuse CR proton flux that effectively escapes from them, and on the productions of secondaries. 
Our main conclusions are that only CR protons with energies $E \lesssim 10^{16}$\,eV are confined in 
galaxy clusters and only produce a negligible amount of secondary \gr\ and neutrinos. Therefore, 
the diffuse CR proton flux above $\sim 10^{16}$\,eV, as well as the diffuse \gr\ and neutrino fluxes 
we computed previously (see e.g. Fig.~\ref{fig:spectrum}) are not affected.

Let us first consider the possibility that CRs residing in a kpc scale jet 
interact with the interstellar medium of the AGN host galaxy. The energy 
loss time $t_{pp}$ for the typical ISM density $n\sim 1$\,cm$^{-3}$ is about 
$3\times 10^7$\,yr, which is longer than the escape time, see Eq.~(\ref{eq:tesc}), 
even for GeV CRs. Thus CRs in the kpc scale jet escape into the interstellar medium of the source host galaxy, 
rather than release their energy inside the jet.

The density of the intracluster medium spread over the Mpc scale of the 
radio lobes has still lower density $n\sim 10^{-2}-10^{-4}$\,cm$^{-3}$, so that the 
$pp$ energy loss time is comparable to or longer than the age of the Universe. 
CRs residing in the radio lobes then escape into the host galaxy cluster of 
the source, rather than dissipate their energy in the lobes.

Therefore, CRs produced in the AGN jet or in the radio lobes escape in the 
host galaxy and galaxy cluster. For a magnetic field strength of $B\sim 1\ \mu$G, which is 
typical for galaxy clusters within a Mpc region, the escape time of 
the very and ultra-high energy CRs contributing to the extragalactic proton flux at Earth 
(above $\sim 10^{16}$\,eV) is small compared to the age of the Universe. 

Lower energy cosmic rays produced by an UHECR source which operates only a limited time ($\sim 10^8$~yr in the case of radio-loud AGNs) are still found in the cluster long since the UHECR source has ceased to exist. Part of their energy will be released while residing in the host galaxy and galaxy cluster. 
Calculating $E_{\rm esc}$ for the case of the host galaxy cluster, one finds that relativistic particles do not escape for $n \sim 10^{-4}$\,cm$^{-3}$. 
Host galaxies and galaxy clusters are therefore expected to give only a subdominant 
contribution to the diffuse neutrino flux. As discussed above, the diffuse \gr\ flux is less sensitive to the value of $E_{\rm esc}$ and, $pp$ interactions
in the host galaxy may contribute to the IGRB 
depending on the slope $\alpha_p$.

\section{Conclusions}
\label{Conclusion}

We have found that star-forming galaxies (normal spiral galaxies, starburst 
galaxies and SF-AGNs) give only a sub-dominant contribution to the high-energy 
CR flux ($E \lsim 10^{18}$\,eV). Both their overall luminosity, their 
redshift evolution which peaks at $z\sim 2-3$, and their maximal energy 
disfavor this source class as the main source of extragalactic CRs up 
to the ankle. The resulting
secondary fluxes are more uncertain, since they depend on the not well
determined grammage CRs cross in their host galaxies and galaxy clusters.
Even keeping the grammage as a free parameter, it is not possible to
explain at the same time a large contribution to the IGRB and
to the IceCube neutrinos or  to the extragalactic part of the IceCube 
neutrino signal. 

In contrast, we have shown that the BL Lac/FR I population as a source for 
extragalactic CRs can explain in a unified way both the observations of 
primary and secondary fluxes. The main reason for this difference with star-forming galaxies or other sources is that  
the number density of BL Lac and FR I galaxies peaks at low redshift, 
$z\lesssim 1$.

More precisely, we found that the extragalactic CR proton flux can be 
explained for any acceleration site (close to the black hole, in the jets or in the lobes) of the BL Lacs/FR I galaxies. 
However, only acceleration close to the black hole (especially at the polar caps, see Section~\ref{CRinterac}) satisfies the required conditions to produce secondary 
\gr\ and neutrino fluxes that can explain the extragalactic IceCube neutrino flux and of the IGRB. For a spectral slope of CR protons close to $\alpha_p=2.2$, 
as suggested by shock acceleration and the escape model, we find that 
such UHECR sources provide the dominant fraction of both the 
isotropic \gr\ background and of the extragalactic part of the astrophysical 
neutrino signal observed by IceCube.

We showed that the difference in the slopes of the proton and the
neutrino fluxes can be explained by the diffusion of primary protons 
in the turbulent magnetic field of CR sources. In the case of Kolmogorov 
turbulence, the power-law of the secondary neutrino spectrum is changed by 
$1/3$, what explains the relatively soft neutrino spectrum with 
$\alpha_\nu\simeq 2.5$ observed by IceCube using a proton spectrum with 
$\alpha_p\simeq 2.1-2.2$.  This mechanism is universal and does not 
depend on the type of sources.

\acknowledgments

GG acknowledges funding from the European Research Council under the 
European Community's Seventh Framework Programme (FP7/$2007-2013$)/ERC 
grant agreement no. 247039.
The work of DS was supported in part by the grant RFBR 
\# 13-02-12175-ofi-m.
The numerical calculations have been performed on the computer cluster of
the  Theoretical Physics Division of the Institute for Nuclear Research
of the Russian Academy of Sciences with support by the Russian Science 
Foundation, grant 14-12-01340.



\end{document}